\documentstyle[aps,prb,graphicx]{revtex}

\begin{document}
\title{Switching behavior of semiconducting carbon nanotubes
under an external electric field}

\author{Alain Rochefort}
\address{Centre de recherche en calcul appliqu\'e (CERCA), Groupe
Nanostructures, Montr\'eal, (Qu\'e), H3X 2H9 Canada 
{\rm and} INRS - \'Energie et Mat\'eriaux, Varennes, (Qu\'e), J3X 1S2 Canada}

\author{Massimiliano Di Ventra}
\address{Department of Physics and Center for Self-Assembled Nanostructures 
and Devices, Virginia Polytechnic Institute 
and State University, Blacksburg, VA 24061-0435}

\author{Phaedon Avouris}
\address{IBM Research Division, T.J. Watson Research Center, 
Yorktown Heights, NY 10598}
\maketitle

\begin{abstract}
We investigate theoretically the switching characteristics of semiconducting
carbon  nanotubes connected to gold electrodes under an external (gate)
electric field. We find that the external introduction of holes
is necessary to account for the experimental observations. We identify
metal-induced-gap states (MIGS) at the contacts and find that the MIGS of an undoped tube
would not significantly affect the switching behavior, even for very  short
tube lengths. We also explore the miniaturization limits of nanotube
transistors, and, on the basis of their switching ratio, we conclude that
transistors with channels as short as 50~{\AA} would have adequate
switching characteristics. 
\end{abstract}

One of the most important findings of carbon nanotubes (NTs) research is that
the current in semiconducting tubes can be switched by an external electric
field which allows semiconducting tubes to be used as the channel of
field-effect transistors (FETs). Several groups have demonstrated such
functional FETs.~\cite{tans,martel,zhou99} The current through the NT has been
found to be maximized at a negative gate bias, indicating that the NTs behave
as {\it p}-type semiconductors, despite the fact that the tubes have not been
intentionally doped. The detailed shape of the source-drain current vs. gate
voltage curves varies somewhat from experiment to experiment, and switching
occurs over a relatively broad voltage range due to complications arising from
charged traps and ``mobile charges" in the gate dielectric.~\cite{sze} As for
the origin of the {\it p}-type character, a number of interpretations have been
proposed, including interaction with the metal electrodes,~\cite{tans} 
impurities and defects introduced during synthesis or processing,~\cite{martel}
or interaction with atmospheric oxygen.~\cite{collins00}

So far, the NT segments used in the FET experiments have been relatively long,
in the order of several hundred nm to a micron. However, the process of device
miniaturization aims primarily at reducing the channel length.  The question
then arises if short nanotube devices can preserve the operational
characteristics of long devices. 

Here we use Green's function techniques to theoretically model nanotube FETs
and address some of these issues. Specifically, we examine the switching
characteristics of both intrinsic and {\it p-} and {\it n}-doped nanotubes of
different lengths. The interaction of the nanotubes with the metal (gold)
electrodes that generates metal-induced gap states (MIGS) is studied, along
with the role of these states in the switching process, and the extent of the
resulting ``metallization" of the short tubes. Finally, the issue of how short
NT FETs can be made is addressed.

Our calculations are performed on segments of semiconducting $(10,0)$
nanotubes. The tubes are bonded with their dangling bond bearing ends to two
gold electrodes.~\cite{rochefort_prb} Each electrode is modeled by a layer of
22 gold atoms in a (111) crystalline arrangement. We also performed
calculations with the tubes lying parallel to the gold electrodes. The main
difference between devices with side-bonding from those with end-bonding is a
higher contact resistance (several M$\Omega$'s as compared to a few 
k$\Omega$'s).~\cite{rochefort_prb,hansson} The electrical transport properties
of the devices were calculated within the Landauer-B\"uttiker formalism using
a Green's function approach.~\cite{datta,economou} The Green's function
${\mathcal{G}}_{NT}$ of the combined electrode-nanotube-electrode system is
written in the form: 
\begin{equation}  
{\mathcal{G}}_{NT}=\big[E{\mathcal S}_{NT}-{\mathcal
H}_{NT} -\Sigma_1-\Sigma_2 \big]^{-1}   
\end{equation}   
\noindent 
where ${\mathcal S}_{NT}$ and ${\mathcal H}_{NT}$ are the overlap and the
Hamiltonian matrices, respectively, and $\Sigma_{1,2}$ are self-energy terms
that describe the effect of the electronic structure of the two leads. The
Hamiltonian ${\mathcal H}_{NT}$ and overlap matrices are determined using the
extended H\"uckel model with $s,p_x,p_y,p_z$ orbitals for each  carbon atom and
one $s$ orbital for each gold atom ($s$ orbitals dominate the DOS of gold near
E$_F$). The field of the gate is approximated as a capacitor field normal to
the NT axis. In this case the Hamiltonian ${\mathcal H}_{NT}$ can be written
as:
\begin{equation}  
{\mathcal H}_{NT} = {\mathcal H}_{NT}^{0} +  {\mathcal H}_{NT}^{1}
\end{equation}   
\noindent
where ${\mathcal H}_{NT}^{0}$ is the Hamiltonian of the unperturbed NT and
${\mathcal H}_{NT}^{1} = e \vec{R}\varepsilon$. Here $e$ is the electron
charge, $\vec{R}$ is the  position of the atoms and $\varepsilon(\vec{r})$ is
the external field.~\cite{prec1}  Only diagonal elements of the Hamiltonian 
${\mathcal H}_{NT}^{1}$ on the extended H\"uckel basis set are taken into
account.  This approximation assumes that all the charge of the carbon atoms is
located at the center of the atoms.~\cite{pablo} This approximation has been
proven to give quite accurate results for semiconductors,~\cite{pablo} and to
reproduce the trends from first-principles transport calculations on molecular
species.~\cite{apl} 
Finally, the transmission function $T(E)$ is given by \cite{datta}:
\begin{equation} 
T(E)=T_{21}=\textrm{Tr} [\Gamma_2 {\mathcal{G}}_{NT} \Gamma_1 {\mathcal{G}}_{NT}^\dagger ],  
\end{equation} 
where $\Gamma_{1,2}=i(\Sigma_{1,2}-\Sigma^\dagger_{1,2})$.  
 
Since in the experiments the potential difference between the source and drain
electrodes is usually very small (typically less than 100 meV),  we evaluate
the conductance at the Fermi energy, $G(E_F)=(2e^2T(E_F))/h$. 

The main panel in Figure 1 shows the variation of the conductance of a
100~{\AA} long $(10,0)$ NT as a function of the applied gate field. Figure 2a
and 2b show the corresponding total density of states (DOS) and the DOS
projected on the $p_x$ and $p_y$ components, for positive and negative gate
fields, respectively. The gap of the NT at zero gate field is about 0.8 eV as
defined by the two peaks at about -0.4 eV (valence-band edge) and +0.4 eV
(conduction-band edge). The {\it n}- ({\it p-}) doping in Figure 1 corresponds
to the conductance calculated with the Fermi level located few meV above
(below) the conduction- (valence-)band edge. In the undoped case the Fermi
level is placed in the middle of the gap. The ballistic limit for a 
semiconducting tube ($T(E_F) = 1$) is also indicated by the horizontal dot
line. By increasing the gate field in the mV/{\AA} range, the C$(p_x)$ and
C$(p_y)$ derived orbitals are shifted by the field to higher (lower) energy for
negative (positive) fields (see Fig. 2). Therefore, the Fermi level  crosses
states with higher (lower) DOS for the {\it n}- ({\it p-})doped NT, with
corresponding increase (decrease) in the conductance. The conductance saturates
when it approaches the  ballistic regime.~\cite{prec} The modulation of the
current (On/Off ratio) achieved with this NT is more than six orders of
magnitude.  Experimentally, FETs with side-bonding configurations have shown
On/Off ratios of $10^5$.~\cite{tans,martel} Recent experiments with end-bonded
low contact resistance tubes have yielded a ratio of $10^6$.~\cite{martel00}

In undoped nanotubes,  the DOS is quite symmetric around the Fermi level and
this should lead to a symmetric conductance vs. gate field curve. However,
figure 1 shows a small asymmetry for small fields. This asymmetry is due to the
metal-induced gap states (MIGS) produced by the interaction of the tube with
the metal electrodes. This interaction produces weak DOS features at about
$-0.15$ eV which extend up to about 8~{\AA} away from the contacts. The
computed MIGS features correlate very well with gap features observed in the
scanning tunneling spectroscopy (STS) of semiconducting NTs on a gold
substrate~\cite{gap1}, whose origin was not previously clarified. 

It is important to explore the possible role of the above gap states in  the
current modulation process. In an undoped NT, the Fermi level ($E_F$) will be
pinned close to the energy position of these states. Furthermore, it has been
shown  experimentally~\cite{odom00} that the band-gap of semiconducting tubes
does not change significantly down to at least 50~{\AA} segments. Figure 3
shows the electronic DOS near $E_F$ of 30~{\AA}, 50~{\AA} and 100~{\AA} long
NTs connected to gold electrodes. Carbon atoms bonded to gold (C-Au)) and gold
atoms (Au) provide the main contributions to the gap states. Assuming that
small variations can occur in the  filling of the MIGS, we have calculated the
conductance by varying the Fermi level across the MIGS energy position. We
found that neither the switching behavior of the NT, nor the conductance at
zero gate field are strongly affected by the movement of $E_F$ (see inset of
Fig. 3). However, a small asymmetry is induced in the switching curve where a
negatively biased gate becomes less effective in switching. This is because, in
this case, $E_F$ is scanned through the MIGS which are localized in the vicinity
of the contacts, giving a lower transmission probability.  The ``metallization" of an
NT segment decreases with its increasing length: the ratio of the DOS of MIGS
to the DOS of the band-edges (van Hove singularities) of the tube is reduced by
a factor of $2/3$ upon doubling the tube length (see Fig.3). This trend should
be valid for other quantum wires, and is in agreement with recent
first-principles calculations on Si nanowires.~\cite{landman}  We thus conclude
that MIGS do not play a significant role in the operation of NT field-effect
transistors.

Next we consider the miniaturization limit of nanotube FETs. Our calculations
and STS measurements~\cite{odom00} indicate that there  are no observable
finite size (length) effects on the band-gap of semiconducting NTs down to at
least 50~{\AA}. Furthermore, the shift of the energy states of the NTs due to
the gate field is independent of the length of the tube because the gate field
is perpendicular to the NT axis. As a result, the conductance of the NTs at
high gate fields should saturate at approximately the same value, irrespective
of the tube's length. Thus, the tube length dependence of the On/Off current
ratio is determined by the Off current. Using this ratio as a  measure of how
well the FET functions, taking MIGS into account and assuming that a ratio of
$10^4$ is the lowest acceptable for practical applications,  then we can see
from Fig. 1c that tubes as short as 50~{\AA} can be used as channels of an FET.
On this scale transport inside the tube is ballistic, there is no energy
dissipation except at the contacts, and THz operation may be possible. These
conclusions are not affected by temperature,  since, for small source-drain
voltages ($\leq$ 0.1V) the Off current is proportional to the  conductance via
a factor eV/4kT, which, at room temperature, is of order one.

In conclusion, we have calculated the switching behavior of intrinsic and {\it
n}- and {\it p}-doped semiconducting nanotubes. We have identified the
metal-induced gap states and examined their role in the field-induced switching
of the tubes.  On the basis of the calculated current modulation, we conclude
that NT segments as small as 50~{\AA} can produce functional transistors.

%
%

\newpage
{\centerline {\bf Figure Captions}}

\vskip1.0cm

\noindent
FIGURE 1. Variation of the transmission of a 100 {\AA} (main panel) and
a 50 {\AA} (upper-left panel) long $(10,0)$ semiconducting nanotube (NT)
as a function of the applied field. Squares: undoped
NT, open circles: $p$-doped, closed-circles : $n$-doped NT. The
On/Off switching ratio is plotted as a function of the NT length
in the upper-right panel.\\

\noindent
FIGURE 2. Effect of a) positive and b) negative gate fields on the 
position of the valence and conduction bands of a 100 {\AA} (10,0) NT. 
The total DOS and projected $p_x$ and $p_y$ DOS are plotted as a function of the 
magnitude of the electric field.\\

\noindent
FIGURE 3. Energy, composition, and relative intensity
of the metal-induced gap states (MIGS) in the vicinity
of Fermi energy for 
a 30~{\AA} (lower panel), 50~{\AA} (middle panel) and a 100~{\AA} (upper panel) 
long (10,0) NT (the valence and conduction band edges are
indicated by the down arrows).
The inset shows the effect of displacing the Fermi level
(full line) to the energy 
of the MIGS (dashed line) on the conductance of the 50~{\AA} NT.

\newpage
\begin{figure}[p]
\vspace*{2.0cm}
\centerline{\includegraphics[angle=0,width=12cm]{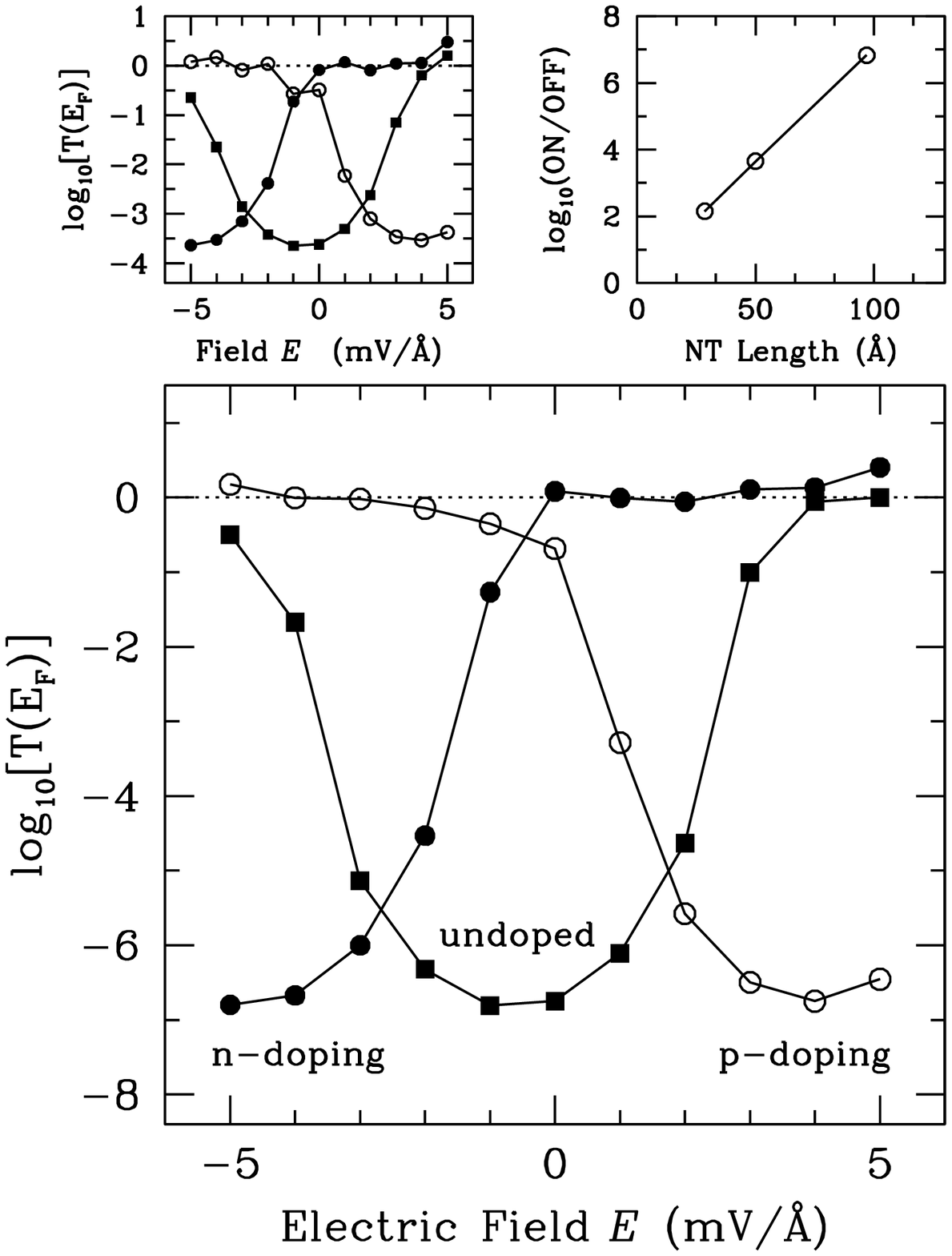}}
\vspace{2.0cm}
FIG.1 Rochefort, Di Ventra, and Avouris
\end{figure}
\thispagestyle{empty}

\newpage
\begin{figure}[p]
\vspace*{1.0cm}
\centerline{\includegraphics[angle=0,width=12cm]{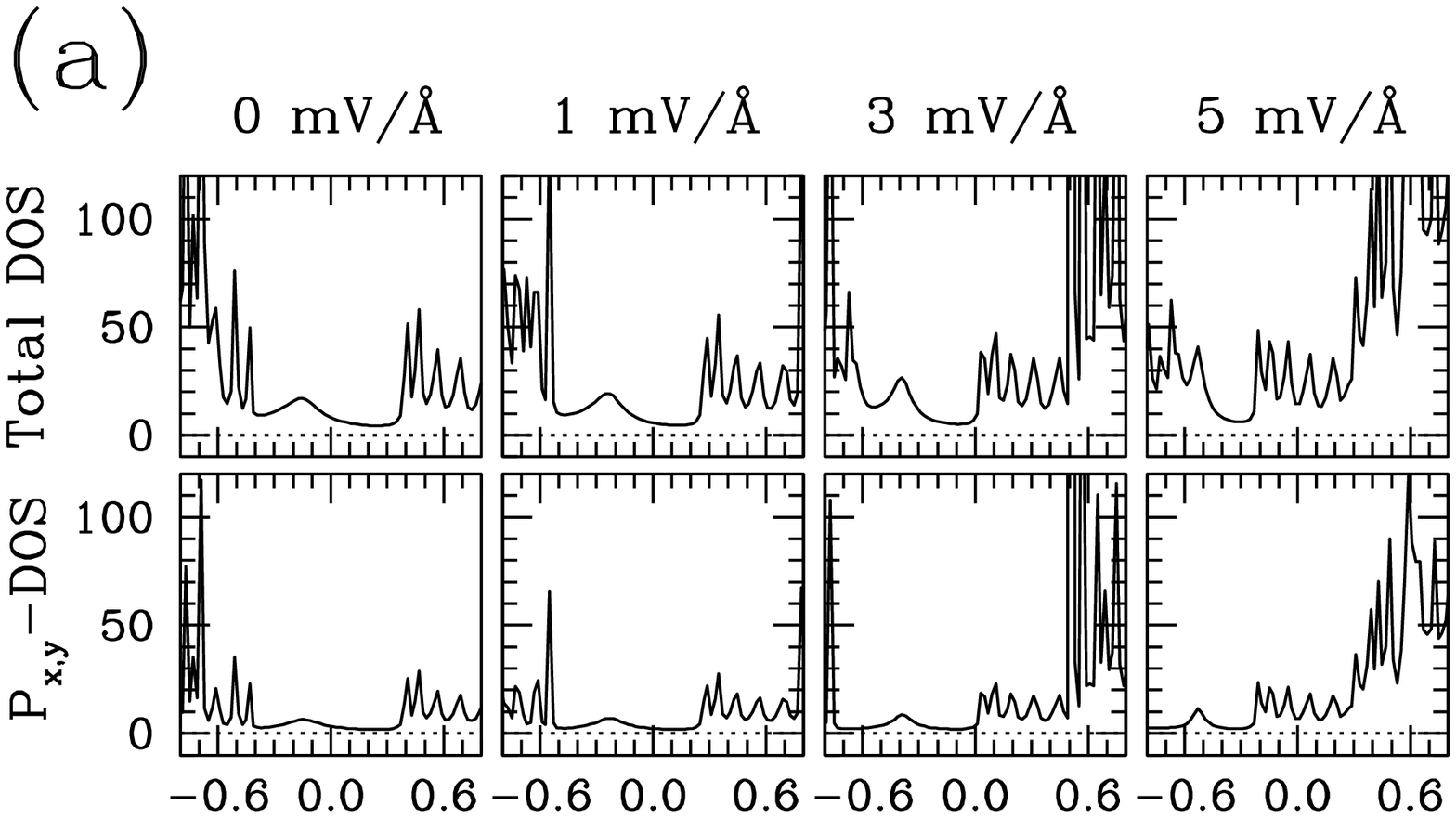}}
\vspace*{0.5cm}
\centerline{\includegraphics[angle=0,width=12cm]{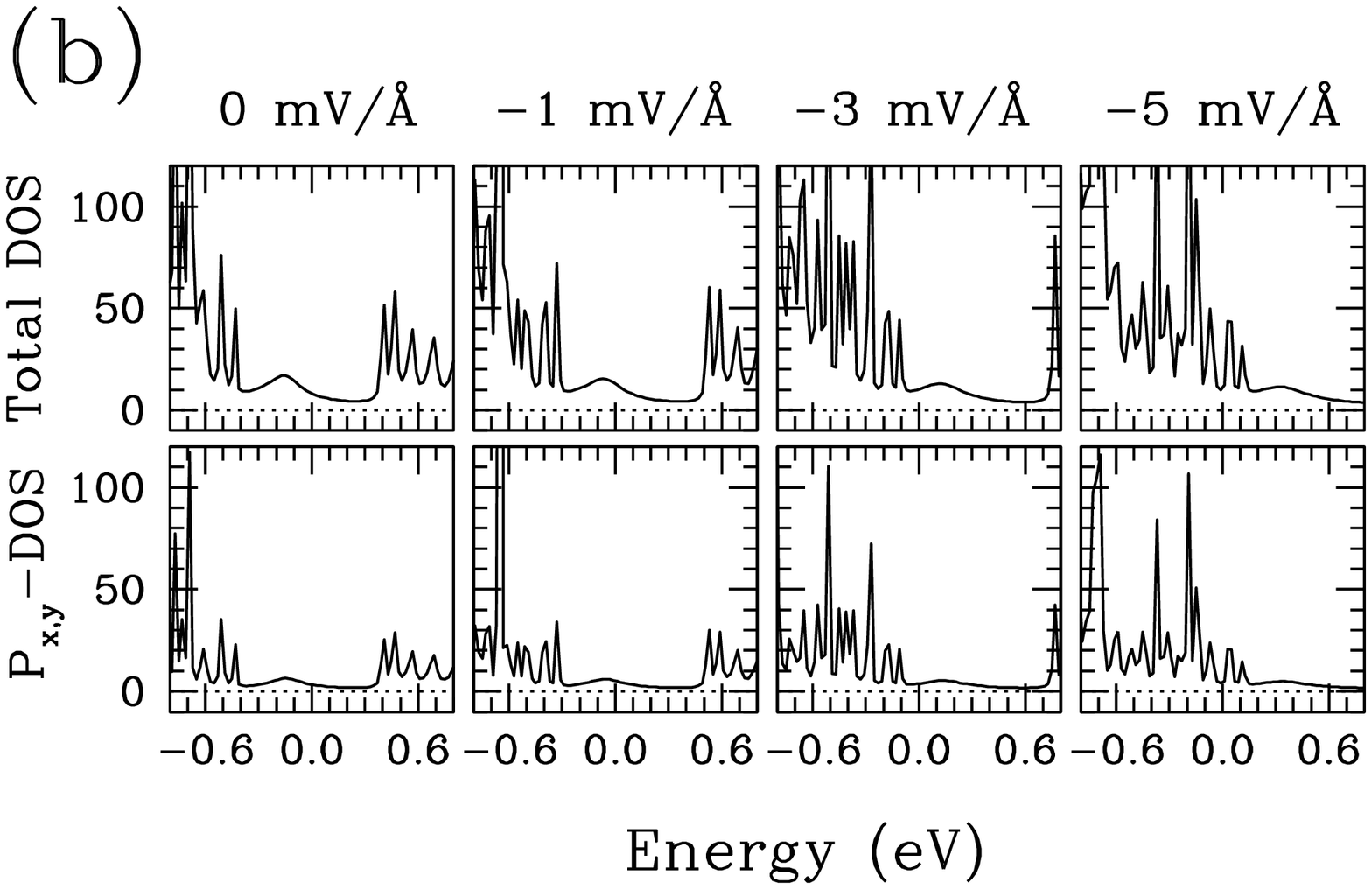}}
\vspace*{2.0cm}
FIG.2 Rochefort, Di Ventra, and Avouris
\end{figure}
\thispagestyle{empty}

\newpage
\begin{figure}[p]
\vspace*{2.0cm}
\centerline{\includegraphics[angle=0,width=12cm]{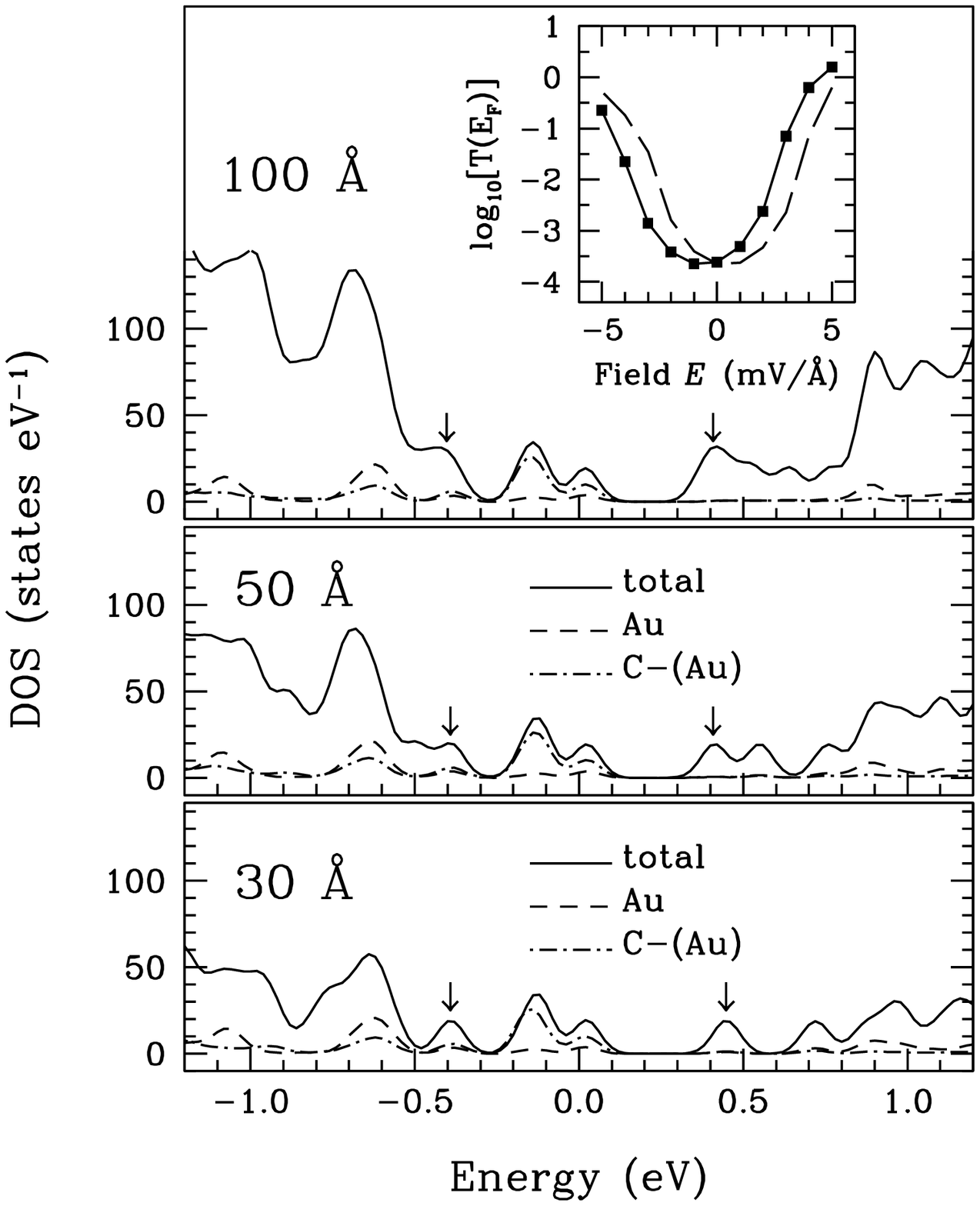}}
\vspace{2.0cm}
FIG.3 Rochefort, Di Ventra, and Avouris
\end{figure}
\thispagestyle{empty}


\begin{references}
\bibitem{prec1} The field is assumed to be non centro-symmetric with respect to the 
tube axis, as in the experiments.
\bibitem{prec} If an extra non-ohmic contact resistance is added to the 
resistance calculated with the present coherent approach, the total 
resistance would saturate to this extra resistance.
\end{references}
\end{document}